# Ultracold hydrogen and deuterium production via Doppler-cooled Feshbach molecules

*Ian C. Lane*

School of Chemistry and Chemical Engineering, Queen's University Belfast, Stranmillis Road, Belfast, UK

BT9 5AG

i.lane@qub.ac.uk

A counterintuitive scheme to produce ultracold hydrogen via fragmentation of laser cooled diatomic hydrides is presented where the final atomic H temperature is inversely proportional to the mass of the molecular parent. In addition, the critical density for formation of a Bose-Einstein Condensate (BEC) at a fixed temperature is reduced by a factor $\left(\frac{m_H}{m_{MH}}\right)^{3/2}$ over directly cooled hydrogen atoms. The narrow Feshbach resonances between a $^1S_0$ atom and hydrogen are well suited to a tiny center of mass energy release necessary during fragmentation. With the support of *ab initio* quantum chemistry, it is demonstrated that BaH is an ideal diatomic precursor that can be laser cooled to a Doppler temperature of ~37 μK with just two rovibronic transitions, the simplest molecular cooling scheme identified to date. Preparation of a hydrogen atom gas below the critical BEC temperature Tc is feasible with present cooling technology, with optical pulse control of the condensation process.

The hydrogen atom is the most abundant element in the Universe and arguably the most important atomic system for testing our current understanding of the physical universe. Naturally, great effort has been expended in laser-cooling[1] and trapping this unique system but, with the lowest energy transition in the vacuum UV (VUV), the recoil temperature $T_r$ is ultimately high[2] and the required laser technology is currently unavailable. In 1998, a Bose Einstein Condensate (BEC) of hydrogen was achieved in a cryogenic (He) environment[3], but the final atom density was very high and the optical access was limited[4]. Furthermore, this technique has thus far not produced a degenerate fermionic gas of deuterium. In this letter a method to produce ultracold H or D atoms via the fragmentation of laser-cooled hydrides and deuterides is outlined. In particular, the benefits of a counter-intuitive scheme, whereby the lowest possible H or D temperatures are produced by pairing with a significantly heavier element, are discussed.

Previously a method has been proposed to produce ultracold samples of carbon[5] and fluorine[6], two elements that are currently impossible to cool to sub-mK temperatures, by the photodissociation of laser cooled diatomics (CH and BeF respectively). The present application of this technique (Fig. 1) significantly extends the method by creating an additional cooling effect, for the final velocity distributions of the atomic fragments are described by lower temperatures than the initial molecular gas. The molecules suitable for the present procedure must satisfy four basic requirements, namely (1) they can be cooled, preferably by laser cooling into the ultracold; (2) they possess very narrow Feshbach resonances to minimize the energy release in the photo-induced fragmentation step; (3) they have a large molecular mass to maximize the cooling effect and finally (4) there exists an efficient pathway between the initial quantum state and the final Feshbach resonance.

The most effective method to fulfill the first condition is direct Doppler cooling with lasers[7], but this requires a diagonal electronic transition and an excited state that does not suffer significant losses, either via radiative decay to alternate electronic states or by predissociation.[5] Stark[8] and Zeeman[9] deceleration of molecules can achieve temperatures into the mK range but not into the ultracold (<1 mK). It is advantageous to use a parent molecule that can be Doppler cooled in order to ensure as low an initial gas temperature as possible, but the technique is not necessarily limited to such molecules.

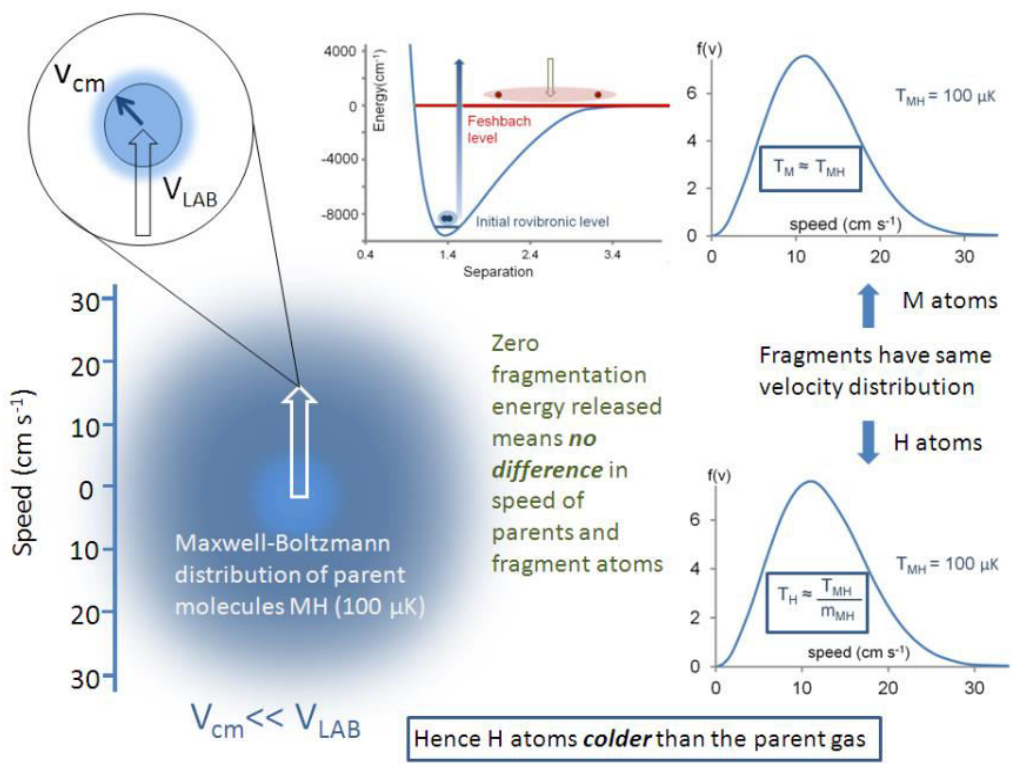

FIG. 1: The dynamics of zero energy fragmentation in an ultracold gas. For molecules moving with the same velocity, the center-of-mass velocity of the fragments is assumed to be considerably smaller than the lab frame velocity of the parent $v_{lab} = v_{rms}$ (*left panel*). This is achieved through excitation via a Feshbach level (*upper middle panel*). The atomic fragments are consequently moving in the lab frame with the same velocity as the parent molecule. A gas of fragmentation products at temperature T will have a Maxwell Boltzmann distribution of fragment speeds. However, the laboratory temperature that describes the fragment velocity distribution depends on the ratio of their mass to the mass of the parent (*right panel*). An extreme example is the production of H fragments from a hydride, such as BaH. A very cold gas (here T = 100μK) will obey the same dynamics, though the low energy fragmentation condition is more demanding to fulfill.

Minimization of the energy released in the fragmentation step with respect to the thermal kinetic energy of the molecules forms the second condition. The initial stage to achieve this is to transfer the molecular population coherently from the final vibronic state in the cooling cycle into a Feshbach resonance, the reverse of the often used magneto-association technique[10] for producing cold molecules from laser-cooled atoms. The important thing to note is that the energy scale associated with the magnetic Feshbach resonances[11] is orders of magnitude smaller than those found in ordinary thermally driven or photochemical bond breaking as such processes involve energy releases equivalent to tens or even hundreds of kJ mol$^{-1}$. By contrast, the relevant energy scale in the ultracold region is 1 μK ≈ 23 kHz ≈ 8.34 μJ mol$^{-1}$. Typically, Feshbach resonance widths are defined by the magnetic field change $\frac{dE}{dB} \approx \Delta\mu B$ and when only the H atoms in the molecule possess a magnetic dipole, the relevant energy scale becomes $1\mu K \approx 14\ mG$.

Thirdly, it is advantageous to include a heavy atomic partner with the hydrogen atom. By using the narrow Feshbach resonances, the center of mass velocities $v_{cm}$ of the fragments are significantly smaller than the initial velocity $v_{MH}$ of the parent molecule and consequently the lab frame fragment velocity becomes $v_M = v_{MH} + v_{cm} \approx v_{MH}$. This condition is particularly difficult to fulfill for the lighter atom in a heteronuclear diatomic and naturally most severe for a hydride. However, for Feshbach widths of $< 100\ mG$ this condition can still be satisfied for hydrogen products from a molecular gas at μK temperatures. Curiously, instead of concern about the energy release in the lighter fragment, when $v_{cm} \ll v_{MH}$ it is possible to exploit a large disparity in atomic mass to create a very cold gas of hydrogen as although the final lab frame velocities are $v_H \approx v_M \approx v_{MH}$, their momenta and associated de Broglie wavelengths are significantly different: $p_H = \frac{m_H}{m_{MH}} p_{MH} \approx \frac{1}{m_{MH}} p_{MH}$. Consequently, the more massive the non-hydrogen atom (and hence the parent), the greater is the subsequent reduction in the hydrogen atom momentum compared to the parent.

The fourth and final condition requires an electronic energy structure that facilitates the efficient transfer of the molecular population from a deeply bound quantum state (the final rovibrational state following laser cooling) into a loosely bound Feshbach state prior to magneto-dissociation. Stimulated Rapid Adiabatic Passage[12] (STIRAP), a fully coherent optical procedure for population transfer, has been used previously for the *formation* of ultracold molecules[10] from cold atoms. An excited molecular electronic state is required to act as a bridge between the initial and target quantum states, a difficult condition to fulfill sometimes as evidenced by recent work on producing ultracold fluorine atoms from BeF[6], where large transition energies and low lying Rydberg states can complicate the available pathways.

The heavier hydrides of alkaline earth metals are potential candidates that fulfill all four requirements and therefore open the door to the production of an ultracold gas of hydrogen or its isotopomer deuterium, the latter of which has yet to be cooled to degeneracy. The lightest examples, BeH and MgH, can be eliminated as they are both prone to predissociation[5] on the most likely cooling transitions, $A^2\Pi \leftarrow X^2\Sigma^+$ and $B^2\Sigma^+ \leftarrow X^2\Sigma^+$. The situation is rather unclear for CaH, but recent *ab initio*[13] and experimental work strongly suggests that only the lowest vibrational level of the $A^2\Pi$ state does not undergo predissociation, and so CaH will not be considered here. While SrH does appear to be a suitable candidate, the superior mass of BaH means this diatomic is the basis of the present study. In previous work we have shown[5,14] that *ab initio* quantum chemistry can be used to identify possible cooling schemes for cooling diatomics. The cc-pV5Z (V5Z) basis set is used for H and all calculations were performed using the MOLPRO suite of programs[26]. The electronic states are found using Complete Active Space SCF (CASSCF) wavefunctions[27] based on the valence orbital space $5a_1$, $2b_1$, $2b_2$ and $2a_2$ (5222) with no core orbitals. Dynamic electron correlation was calculated by Multi-Reference Configuration Interaction (MRCI)[28], using the relaxed Davidson correction[29] throughout. These calculations were augmented with additional points using the larger active spaces 6532 (for the Rydberg component of the Π states) and 7432 (for the $\Sigma^+$ Rydberg states). Only states of doublet multiplicity were calculated, and since all the $^2\Sigma^-$ valence states are repulsive they are not discussed further. The ab initio data points were fitted using the LEVEL program (version 8.0)[30] to obtain the bound rovibronic levels, spectroscopic constants (table 1), FC factors and radiative lifetimes. In the present work, there is a wealth of experimental data too and this information is combined with high level *ab initio* calculations to develop as accurate a set of potential curves as possible for BaH and to obtain the bound rovibronic levels, spectroscopic constants, Franck Condon (FC) factors and radiative lifetimes. The ground $X^2\Sigma^+$ state dispersion $C_6$ co-efficient was computed as 147.5(4) using the polarizibility data of Derevianko *et al*[15], and the excited state values were estimated for the purposes of fitting the potentials. The lifetime of the BaH $B^2\Sigma^+$ v = 0 level has been measured at 125 ± 2 ns (31 μK)[16] and the calculated transition dipole was scaled to match the experimental result, which allowed us to determine the corresponding lifetime of the $A^2\Pi$ v = 0 level as 105 ns and a corresponding Doppler temperature of 37 μK. It is worth noting that the recoil temperature for the $A^2\Pi$ v = 0 ↔ $X^2\Sigma^+$ v = 0 transition is just 125 nK, thanks to the long cooling wavelength and large molecular mass.

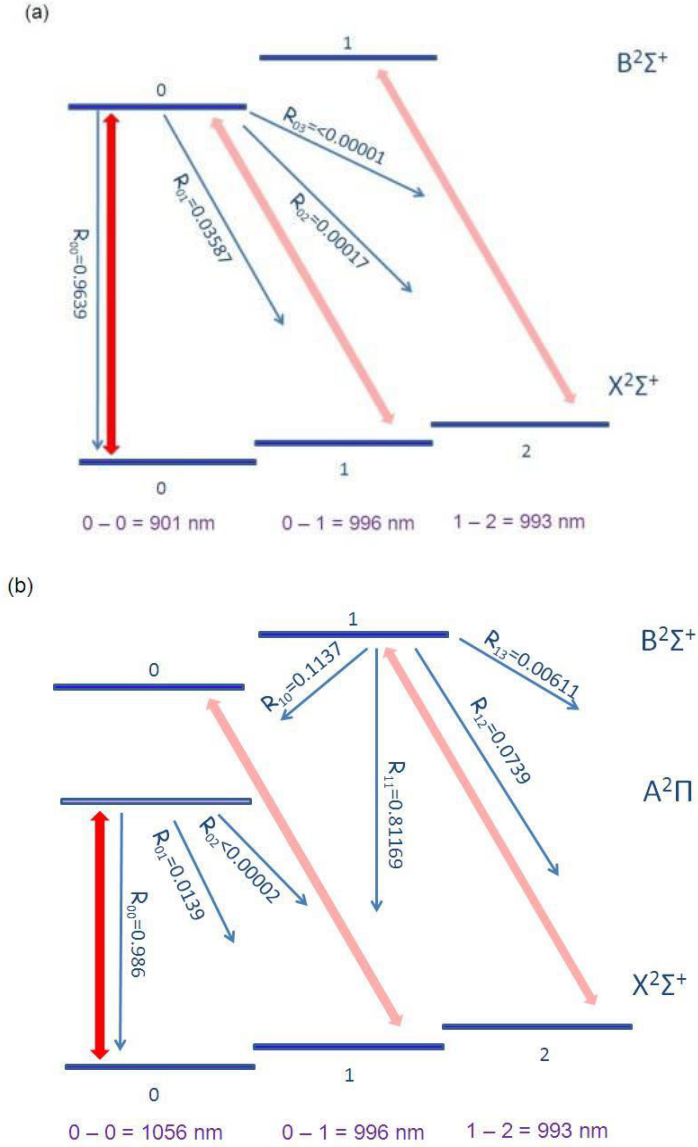

FIG. 2: Alternative BaH cooling schemes. The relative strengths of the photon decay pathways are labeled $R_{v'v''} = \frac{A_{v'v''}}{\sum A_{v'v''}}$ rather than as FC factors. Cooling wavelengths indicated at the bottom.

With similar cooling rates available via both the $A^2\Pi$ and $B^2\Sigma^+$ states, two cooling schemes (Fig. 2) are proposed, though the crucial factor for successful cooling is the minimization of any radiative losses. Consequently, we determined the ratio of the loss channels (a) non-diagonal $A^2\Pi_{1/2}$ v = 0 ↔ $X^2\Sigma^+$ v = 1,2 or 3; (b) $B^2\Sigma^+$ v = 0 ↔ $X^2\Sigma^+$ v = 1,2 or 3; (c) $A^2\Pi_{1/2}$ v = 0 ↔ $A'^2\Delta_{3/2}$ v = 0 and (d) $B^2\Sigma^+$ v = 1 ↔ $A^2\Pi_{1/2}$ v = 0, 1 and 2. Computing the relevant FC factors are not enough when there exist more than one lower electronic state. Furthermore, the large vibrational energy separation in hydrides means that even within a single electronic manifold the FC factors fail to capture the large changes in vibrational branching ratios $R_{v'v''}$, determined by the relative Einstein A co-efficients;

$$R_{v'v''} = \frac{A_{v'v''}}{(\sum_{v''} A_{v'v''})} \quad (1)$$

because they do not account for the $\omega^3$ dependence of radiative decay. Radiative decay to higher vibrational levels with their corresponding lower emission frequencies are suppressed compared to the desired cyclic 0-0 transition.

The *ab initio* results reveal that when cooling is primarily conducted on the $A^2\Pi_{1/2}$ v = 0 ↔ $X^2\Sigma^+$ v = 0 transition, only $A^2\Pi_{1/2}$ v = 0 ↔ $X^2\Sigma^+$ v = 1 contributes significantly to the loss. This is helped by the large change in emission wavelength, which affects the rate of decay, of the competing decay channels due to the large vibrational frequency of the ground state molecule. In conclusion, the cooling scheme 2(b) is favored, though a reduced two color laser cooling scheme is perfectly feasible. The smaller vibrational frequency in the deuteride weakens the effect of changes in emission wavelength, and the branching ratio for $A^2\Pi_{1/2}$ v = 0 ↔ $X^2\Sigma^+$ v = 2 is consequently five times higher in BaD than BaH. Thus it is only possible to achieve 8 - 9000 cooling transitions with two laser colors. This wavelength suppression effect is a major advantage of cooling hydrides (and deuterides) over other diatomic molecules. A recent *ab initio* calculation using relativistic potentials (the discrepancies between the theoretical and experimental $r_e$ values < 0.01Å[17]) has computed the bond dissociation energy $D_0(X^2\Sigma^+)$ as 16681 cm$^{-1}$ (2.06 eV) which ensures that the lowest vibrational levels v′ = 0 and 1 of the $B^2\Sigma^+$ state are below the dissociation limit. Therefore, predissociation is not a concern for laser cooling BaH with the scheme proposed.

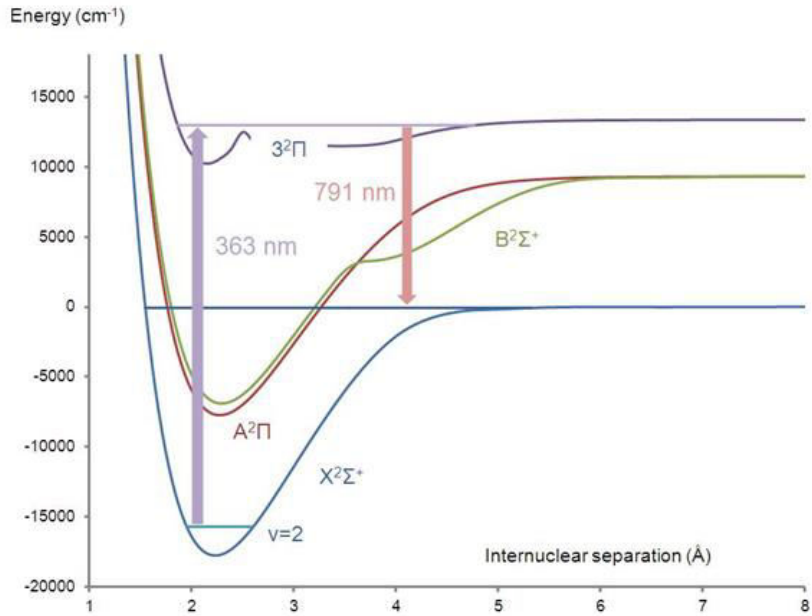

FIG. 3: Proposed STIRAP pathway in BaH and the corresponding pulse wavelengths. Also marked are the bound state potentials involved in the cooling cycle.

BaH has a number of odd and even isotopes but due to the larger abundances the focus here will be $^{137}$BaH and $^{138}$BaH. Even isotopes of the $^1S_0$ atoms are usually without nuclear spin and cannot contribute to the Feshbach linewidth. This is not the case for odd isotopes, which always have half integer spin that can also couple to the molecular spin via $\Delta\zeta_{Ba}(R)$.[18] For the corresponding alkali-alkaline earth diatomics $^7$Li$^{173}$Yb and $^7$Li$^{174}$Yb, the observed widths are typically 100x larger for the $^{173}$Yb species but the resonance width is typically < 5 mG even for the broadest lines. This corresponds to an energy (temperature) range of $\Delta E \approx 7$ kHz (0.16 μK) and makes $^{137}$BaH an attractive candidate for ultracold hydrogen production. To ensure a very narrow[19] width, it is advantageous to have a small background scattering length $a_{bg}$ and to ensure a large difference in the energy shift with magnetic field between the open and closed scattering channels. It is presently only possible to obtain a reliable value for $a_{bg}$ from experimental measurements with ultracold atoms but it has been shown that there is a 75% chance of it lying between $-\bar{a}$ and $3\bar{a}$, where $\bar{a}$ is the mean scattering length which is known[20] to be proportional to $(2\mu C_6)^{1/4}$. Using the earlier dispersion constant, the estimated $a_{bg}$ value for BaH (BaD) lies between -13$a_0$ and 39$a_0$ (-15$a_0$ and 45$a_0$), a comparable range to LiYb. Furthermore, the narrowest (and only) resonances in $^{138}$BaH will be found for $m_{fH}$ = 0, while in $^{138}$BaD it will be $m_{fD}$ = +1/2 (there will be an additional resonance for $m_{fD}$ = -1/2). While Δ(B) scales as B$^2$ for low magnetic fields before leveling out, this effect will be counteracted by the shallower energy slopes at lower fields for $m_{fH}$ = 0 ($m_{fD}$ = +1/2) below 500G (100G).[21]

To ensure that optical transfer is possible, a pathway to the highest bound levels of the X$^2$Σ$^+$ state is found using the *ab initio* potentials. Despite lying thousands of wavenumbers above v = 0, the highest (final) vibrational level close to the dissociation limit will have a significantly longer radiative lifetime[22] (≈ 0.1-1000 s depending on the binding energy) than all the other excited vibrational levels, even the first excited v = 1 state, while the inelastic collision rate constant[23] is typically only of the order 10$^{-11}$ cm$^3$ s$^{-1}$. The present calculations suggest that excitation via the 3$^2$Π state will facilitate the population of the highest vibrational levels of the X$^2$Σ$^+$ state by the STIRAP process, thanks to its double well structure (Fig. 3).

Fig. 4(a) presents the temperature dependence of the final hydrogen (and deuterium) fragments on the mass of the parent molecules. Recently, a nearly uniform atom trap has been developed by Hadzibabic and co-workers for the production of a Rb BEC and an accurate transition temperature of $T_{cRb}$ = 92 ±3 nK measured[24] for $n$ ≈ 3 x 10$^{12}$ atoms cm$^{-3}$. The general dependence of $T_{cm}$ on particle mass $m$ and density $n$ is given by $T_{cm} \propto \frac{n^{2/3}}{m}$. Using this experimental result for Rb, the corresponding $T_{cH}$ and $T_{cD}$ are added to Fig. 4(a) assuming an identical particle density. The H(D) atoms from zero energy photofragmentation of CaH (CaD) and heavier hydrides are at a lower temperature than $T_{cH(D)}$ when the parent temperature is 20 - 100μK. But the most significant benefit of this technique is the large reduction in the critical density required for quantum degeneracy. Fragmentation of the hydride leads to a larger de Broglie wavelength for the hydrogen atoms than the parent molecules $\lambda_H \approx m_M \lambda_M$, increasing the phase space. The mass dependence of critical density is $(mT)^{3/2} = n$, where $n$ is the density of the gas in particles cm$^{-3}$ and $m$ is the mass, so for two species ($m_{MH} > m_H$) at a fixed temperature, the ratio of critical densities is $\frac{n_{MH}}{n_H} = \left(\frac{m_{MH}}{m_H}\right)^{3/2}$. When $m_{MH}$ is a molecule that fragments with zero energy release into fragments, one of which is hydrogen, the latter will have an increased de Broglie wavelength $\left(\frac{m_{MH}}{m_H}\right)$, and thus the ratio of critical densities $n_H$ for an atomic hydrogen gas and hydrogen fragmentation products $n_H'$ is

$$\frac{n_H'}{n_H} = \left(\frac{m_H}{m_{MH}}\right)^3 \left(\frac{m_{MH}}{m_H}\right)^{3/2} = \left(\frac{m_H}{m_{MH}}\right)^{3/2} \quad (2)$$

Fig. 4(b) documents the critical density as a function of the initial temperature for a gas of hydrogen atoms, BaH molecules and for the hydrogen atoms produced by fragmentation of the same BaH gas. The advantages of the fragmentation scheme, that exploits the twin effects of the $\left(\frac{m_H}{m_{MH}}\right)^{3/2}$ mass dependence (at a fixed temperature) in combination with the lower temperatures possible by laser cooling the parent molecule compared to H atoms directly ($T_r$(BaH) << $T_D$(BaH) << $T_r$(H)), leads to a dramatic fall in the hydrogen critical density. This demonstrates that a relatively dilute gas of ultracold H below $T_{cH}$ can be created with benefits such as a longer lifetime and smaller collisional shifts. Although the 37 μK initial temperature (Doppler temperature of A$^2$Π ↔ X$^2$Σ$^+$ laser cooled BaH) is not demanding for an atomic gas, to date the lowest temperature achieved by direct laser cooling of a molecular sample[25] (YO) is around 2 mK. However, this is comparable with $T_r$ for direct laser cooling of atomic hydrogen and with a critical density (139)$^{3/2}$ ~1600 times smaller.

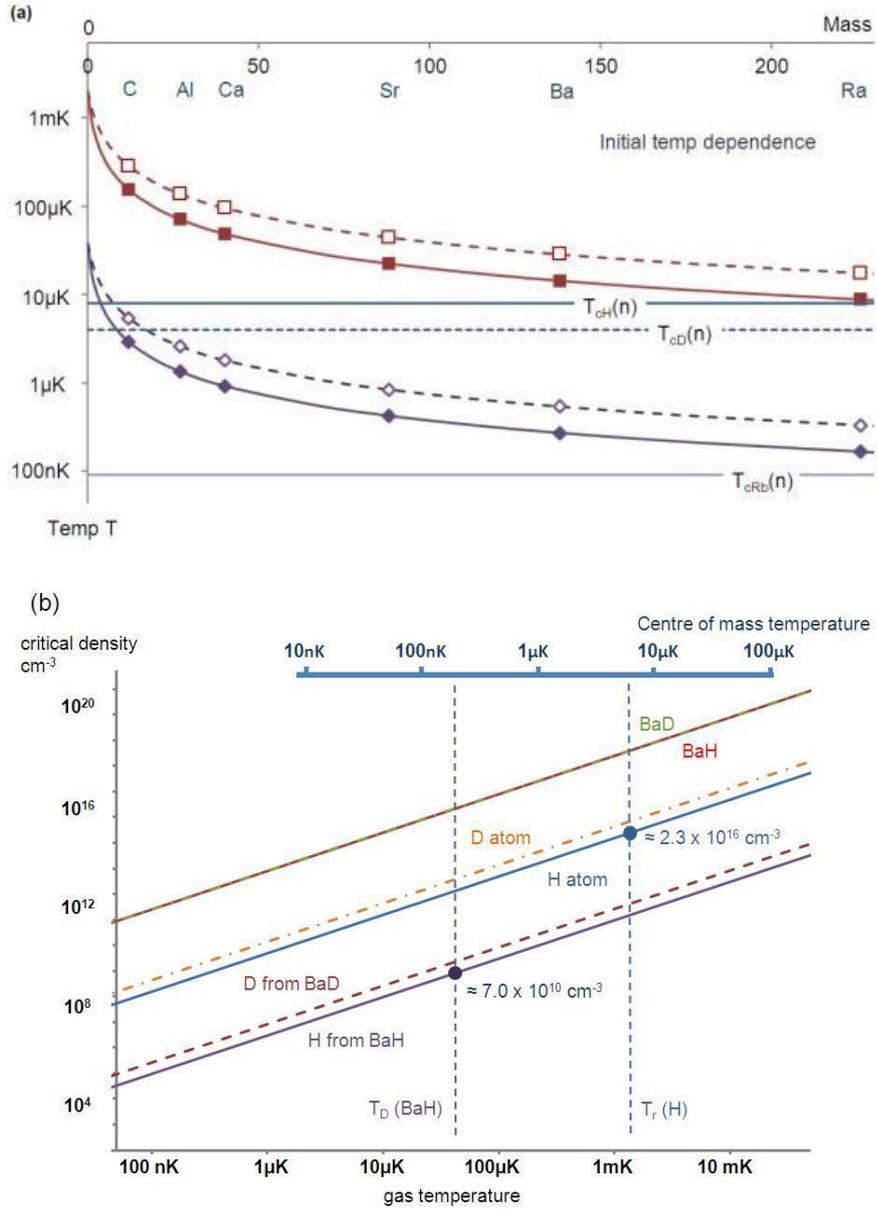

FIG. 4: The effect of parent mass on the final H atom temperatures following molecular fragmentation. (a) For molecules with initial temperature 37μK or 2mK prior to fragmentation (the center-of-mass fragmentation energy must be considerably smaller than the thermal energy of the parents). Dashed lines correspond to the deuterated species. Also marked are the values of $T_c$ for Rb, D and H at the experimental density[24] $n$ used by Hadzibabic and co-workers. (b) The effect of molecular mass on the critical density to achieve a quantum gas, compared with direct cooling of H or D atoms alone. Also marked are the recoil temperature for H atom cooling and the Doppler, recoil temperatures for BaH $A^2\Pi_{1/2} \leftrightarrow X^2\Sigma^+$ cooling and the corresponding critical densities. The center of mass temperature represents the maximum energy that can be released in the fragmentation event while preserving the lab frame velocity distribution. The origin of the significant reduction in critical density is discussed in the text.

An ultracold gas of hydrogen atoms would be the ultimate laboratory to probe fundamental questions about our universe. In this letter a technique to create ultracold hydrogen and, uniquely, deuterium atoms by the fragmentation of ultracold and trapped diatomics has been analyzed. Four important conditions that the molecular parent should fulfill in order to achieve the lowest possible temperatures have been identified. Using *ab initio* molecular potentials and transition dipoles of BaH relevant for the production of ultracold H via fragmentation of laser cooled molecules, the most efficient laser cooling scheme and principal decay channels have been found. A single-color scheme for BaH will ensure 50-70 optical cycles (which should be observable in experiments like Hummon et al[25]) a two-color scheme 40-50 000, while a three-color experiment is effective for > $10^6$ cycles. A STIRAP excitation process via the $3^2\Pi$ state (a mixed Rydberg/valence state) appears feasible with good oscillator strengths. Exploiting the $\frac{m_{H(D)}}{m_{MH(D)}}$ dependence of the final temperature when $v_{cm} \ll v_{MH(D)}$ it is possible to create a dilute gas of hydrogen or deuterium below $T_c$ with full optical access and a reduction of the critical density by a factor of at least 200 000 compared to directly laser cooled hydrogen atoms.